# A Theoretical Investigation of the Possible Detection of $C_{24}$ in Space

SeyedAbdolreza Sadjadi [1, 2], Sun Kwok[2,3], Franco Cataldo[4], D.A. García-Hernández[5,6], and Arturo Manchado[5,6,7]

[1]*School of Physics and Astronomy, Sun Yat-sen University, Zhuhai, China*

[2]*Laboratory for Space Research, The University of Hong Kong, Hong Kong, China*

[3]*Dept. of Earth, Ocean, and Atmospheric Sciences, University of British Columbia, Vancouver, Canada*

[4]*Istituto Nazionale di Astrofisica-Osservatorio Astrofisico di Catania, 95123 Catania, Italy*

[5]*Instituto de Astrofísica de Canarias, E-38205 La Laguna, Tenerife, Spain*

[6]*Universidad de La Laguna, Departamento de Astrofísica, E-38206 La Laguna, Tenerife, Spain*

[7]*Consejo Superior de Investigaciones Científicas, Spain*

Email of corresponding author:  skwok@eoas.ubc.ca

# A Theoretical Investigation of the Possible Detection of $C_{24}$ in Space

Astronomical infrared spectral features at ~6.6, 9.8 and 20 μm have recently been suggested as being due to the planar graphene form of $C_{24}$ carbon cluster. Here we report density functional theory and coupled cluster calculations on wavefunctions stability, relative energies, and infrared spectra of four different types of $C_{24}$ isomers, including the graphene and fullerene forms. The types of vibrational motions under these bands are also discussed. Among the four isomers, we find that the astronomical data are best approximated by the graphene form of $C_{24}$.

**Introduction**

The planar form of $C_{24}$ (graphene sheet) has recently been suggested to be the chemical carrier of three infrared bands at 6.6, 9.8 and 20.1 μm observed in the spectra of planetary nebulae, objects in the late stage of stellar evolution (1, 2). This interpretation is supported by theoretical results of density functional theory (DFT), showing that the infrared transitions of planar $C_{24}$ match the observed band positions, and the fact that this isomer is likely to be more abundant in space since it is energetically more favourable than its fullerene form (3).

However, more accurate calculations at the basis set limit of couple cluster (CCSD(T)) theory showed that these two isomers are equal in energy (4). Assuming the similar kinetic pathways for the formation of these isomers, this new theoretical results implies that both isomers should have similar abundance in astronomical sources and thus both could have vibrational bands detectable in the infrared.

Prompted by these apparent contradictory results, we examined the electronic structures, relative energies and finally the simulated infrared spectra of more $C_{24}$ isomers (Figure 1). These are graphene (Figure 1a), fullerene (Figure 1b), ring (Figure 1c), and bowl (Figure 1d) forms. We hope that these results can help discriminate the isomer or isomers expected to be present in astronomical sources, and thus provide clues on the possible reaction pathways to the formation of such carbon clusters in space.

**Details of quantum chemical models**

Density functional theory (DFT) was applied via RB3LYP formalisms for geometry optimization and normal mode frequency calculations. All isomers were characterized as local minimum. We used UB3LYP for single point calculations at triplet electronic state of these carbon clusters. Scale factors for normal modes vibrational frequencies and zero point vibration are obtained from Laury et al. (5)

In order to accurately calculate the relative energies among the $C_{24}$ isomers, single point couple cluster calculations in form of CCSD(T) were performed to cover the electronic correlation effects. In all calculations we applied the polarization consistent basis set, PC1. Zero point energy corrections to CCSD(T) final energies are obtained from B3LYP calculations.

All DFT calculations were performed by firefly8.1.1 (6) running on HKU supercomputer facility, HPC2015. CCSD(T) calculations were conducted by molpro (7) compiled on Quantum Cube TM.

**Results and discussion**

*Wavefunction character and electronic ground state*

We conducted the frequency calculations on the presumed singlet electronic state via single reference DFT formalism for all isomers. The results of the singlet-triplet energy difference and characteristic of the wavefunction for each isomer are summarized in Table 1. These results show that except for the bowl isomer (Figure 1d), which has a high degree of multi-reference character in its wavefunction, the electronic structure and the associated molecular properties (including the infrared vibrational frequency and intensities of transitions) of all other isomers can be well modelled by present single reference DFT formalism. The large gap between triplet and singlet electronic states in graphene and fullerene isomers, in addition to their wavefunctions single-reference character, are evidence of the singlet ground electronic state for these isomers. The singlet-triplet electronic states crossover and multi-reference character of the wavefunction in two other isomers (ring and bowl) show that the calculated infrared data at presumed singlet state may not accurately reflect the vibrational signature of these isomers, especially for bowl isomer with large triplet state stabilization.

*Relative energies*

The relative energies values for all four $C_{24}$ isomers are presented in Table 2 at both DFT and CCSD(T) levels. The corresponding values estimated at CCSD(T) complete basis set (CBS) limit (4) are presented in the last column. In spite of the fact that the RMSD error associated with DFT relative energies have been estimated to be 8-34 kcal/mol (4), the graphene form is found to be the lowest energy isomer in both DFT and CCSD(T) level of calculations. The next low energy isomer is fullerene form (Table 2). The trend of calculated relative energies among the $C_{24}$ isomers is in good agreement with the values at the CBS limit (4).

### *Simulated emission spectra*

We have shown that the B3LYP/PC1 formalism is a relatively reliable model to explore the electronic structures of graphene, fullerene, and ring forms, whereas the calculated infrared data at singlet state for bowl isomer is less reliable. We also assume that due to small crossover (4.87 kcal/mol) between two electronic states the electronic ground state of ring isomer is singlet and thus the calculated infrared data is still valid. The calculated spectra were broadened with a Drude model with a temperature (*T*) of 500 K and a broadening profile with FWHM=0.03 μm to simulate the astronomical infrared emission spectral data (8). The resultant spectra for the 4 isomers are shown in Figure 2. For comparison, the infrared spectrum of the bowl isomer is also shown together with the other isomers.

Figure 2 shows that the graphene form of $C_{24}$ displays sharp bands at 6.708, 9.847 and 22.480 μm, close to the reported astronomically observed bands at 6.6 μm, 9.8 μm and 20 μm (1). The theoretical line strength ratios are also qualitatively in agreement with the observed line ratios. None of the other isomers show strong bands near these observed features.

### *Final assessment of the band positions*

We evaluated possible errors in our theoretical band positions in by comparing the simulated data for of $C_{60}$ at solid state to the laboratory FTIR data (Figure 3 and Table 3). We observed that the simulated bands below 10 μm show red-shifts in wavelength, while the larger wavelength bands show blue-shifts. From the observed differences seen in $C_{60}$ shown in Table 3, we adjusted the positions of infrared bands of graphene form of $C_{24}$ to 6.599 μm, 9.761 μm and 22.543 μm. The wavelengths of the first two bands agree very well with the astronomical observations.

The 20.1 μm feature observed in astronomical spectra (PNe SMC 24 and LMC 02, ref (1)) is the weakest and less reliable feature from the three $C_{24}$-like features. So, higher resolution and higher S/N data are needed and this could be done with the upcoming *James Webb Space Telescope (JWST)*. Indeed, the two sources showing $C_{24}$-like features, when observed at higher resolution with *Spitzer* (data only available for the 12-20 μm range; see Figure 2 of ref (1)), display a plethora of unidentified features not seen at lower resolution. But the S/N of the high-resolution (*R*~600) *Spitzer* spectra is not very good and the reliability of most of these unidentified features remains uncertain. So, better data on these sources with $C_{24}$-like features may provide a definitive answer about the possible detection of $C_{24}$ (planar vs fullerene or both).

*Vibrational motions in graphene form*

From visual inspections of the animation of vibrational normal modes we concluded that all the above three bands of graphene $C_{24}$ are due to in-plane modes vibrations (Figure 4). The vibration under 6.708 μm can be viewed as C–C stretching mode, 9.847 μm as C–C–C angle bending mode and 22.480 μm as ring distortion mode.

*Fullerene $C_{24}$*

Recently, Bernstein et.al (9) assigned two astronomical infrared bands at 11.2 and 12.7 μm as due to fullerene $C_{24}$, based on the B3LYP/6-31G(d) basis set. We found general agreement between our calculated band positions and relative intensities with those calculated by Bernstein et.al (9), in spite of the differences in the chosen basis set and the scale factors. Specifically, we found the major bands of fullerene $C_{24}$ to occur at 8.076, 9.957, 11.027, 12.304, 15.82, 16.405, 17.241 and 24.212 μm (Figure 2), compared to the respective wavelengths of bands at 7.91, 9.72, 11.01, 12.27, 15.95, 16.86, 18.38 and 19.91 μm of Bernstein et.al (9). The strongest band is at 18.38/17.241 μm.

**Conclusion**

Our results show that both the graphene and fullerene forms of $C_{24}$ have very stable singlet electronic ground states. This is evidenced by their large gap (positive value) between singlet and excited triplet states and their single-reference characters of their electronic wavefunctions. It is anticipated this will provide stability against dissociation from absorption of interstellar ultraviolet/visible radiation, allowing for both graphene and fullerene form to survive in space. The other two isomers, namely ring and bowl, do not exhibit such stability in their electronic structure. If $C_{24}$ exists in space, it is likely that both the graphene and fullerene forms would be present.

Comparing to the astronomical observations, we find that the observed lines are best approximated by the graphene form of $C_{24}$. Further observations with the *JWST* will be needed to confirm this identification.

Given that now fullerene ($C_{60}$) has been detected in circumstellar and interstellar environments (10), it is quite possible that the fullerene form of $C_{24}$ would also be present in similar environments. The suggestion that the carriers of the unidentified infrared emission bands are breakdown products of fullerene (11, 12) could have implications on the formation pathway of $C_{24}$. We hope that the theoretical results in this paper will help confirm the identification of $C_{24}$ in space.

**Acknowledgments**

The Laboratory for Space Research was established by a special grant from the University Development Fund of the University of Hong Kong. This work is also in part supported by a grant to SK from the Natural Sciences and Engineering Research Council of Canada. DAGH, AM, and FC acknowledge support from the State Research Agency (AEI) of the Spanish Ministry of Science, Innovation and Universities (MCIU) and the European Regional Development Fund (FEDER) under grant AYA2017-88254-P.

Table 1. Electronic ground state and multi-reference character of wavefunction of $C_{24}$ isomers.

Table 2. Zero point energy corrected relative energies calculated at B3LYP/PC1 and CCSD(T)/PC1 levels for $C_{24}$ isomers.

Table 3. Comparison between experimental and simulated infrared bands positions at B3LYP/PC1 level for solid state $C_{60}$ at $T$=323.15 K

Figure 1. Local minimum geometries of four isomers of $C_{24}$ calculated at B3lYP/PC1 level. The C atoms are shown in grey

Figure 2. Simulated infrared emission spectra ($T$=500 K) of four isomers of $C_{24}$ at B3LYP/PC1 level

Figure 3. A comparison between solid state FTIR data and combined Drude-B3LYP/PC1 simulation at $T$=323.15 K for $C_{60}$.

Figure 4. Vibrational motions in graphene form of $C_{24}$ at a) 6.708 μm, b) 9.847 μm and c) 22.480 μm. The C atoms are shown in gray. Displacement vectors are in red.

Table 1. Electronic ground state and multi-reference character of wavefunction of $C_{24}$ isomers.

| isomers | $\Delta E_{singlet-triplet,DFT}$ (kcal/mol) | $T1_{CCSD(T)}$[a] | $D1_{CCSD(T)}$[a] | Ground (character) |
|---|---|---|---|---|
| Graphene ($D_{6h}$) | 87.86 | 0.01464 | 0.03867 | singlet (single-ref) |
| Fullerene ($D_6$) | 329.21 | 0.01571 | 0.04859 | singlet (single-ref) |
| Ring ($D_{12h}$) | -4.87 | 0.01415 | 0.03748 | triplet (single-ref) |
| Bowl ($C_1$) | -52.06 | 0.07102 | 0.61888 | triplet (multi-ref) |

a) Criteria for wavefunction to be single reference: T1<0.025 (13) and D1<0.100 (14)

Table 2. Zero point energy corrected relative energies calculated at B3LYP/PC1 and CCSD(T)/PC1 levels for $C_{24}$ isomers.

| isomers | $\Delta E_{rel,DFT}$ (kcal/mol)[a] | $\Delta E_{rel,CCSD(T)}$ (kcal/mol)[b] | $\Delta E_{rel,CCSD(T)}$ (kcal/mol)[c] |
|---|---|---|---|
| Graphene (D6h) | 0.00 | 0.00 | 0.00 |
| Fullerene (D6) | 33.29 | 14.31 | 1.72 |
| Ring (D12h) | 41.01 | 129.51 | 74.58 |
| Bowl (C1) | 185.81 | 95.61 | - |

a) scale factor of 0.9880 is used from ZPVEB3LY P from ref (5).

b) ECCSD(T)+ZPVEB3LYP .

c) ECCSD(T),cbs +ZPVEPBE0.from Manna and Martin (4) , by adding table 5 and table 7 data of this reference.

Table 3. Comparison between experimental and simulated infrared bands positions at B3LYP/PC1 level for solid state $C_{60}$ at $T$=323.15 K

| $\lambda_{exp}$ (μm) | $\lambda_{DFT}$ (μm) | $\lambda_{exp-DFT}$ (μm) |
|---|---|---|
| 7.007 | 7.116 | -0.109 |
| 8.473 | 8.559 | -0.086 |
| 17.401 | 17.364 | 0.037 |
| 19.064 | 19.001 | 0.063 |

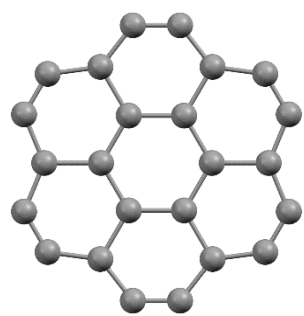 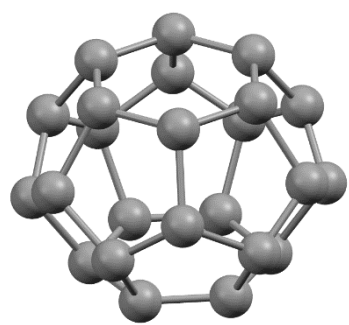

**a)** $D_{6h}$  **b)** $D_6$

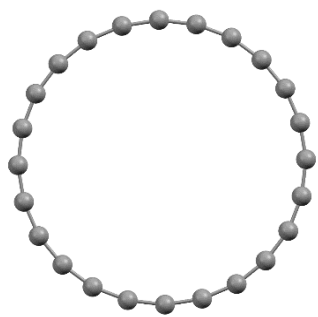 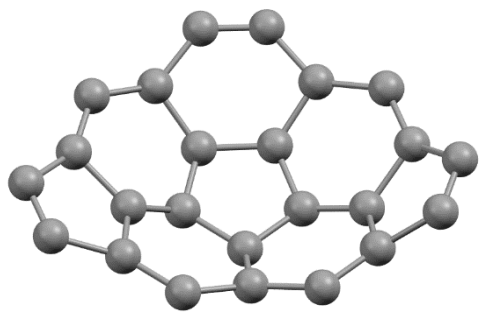

**c)** $D_{12h}$  **d)** $C_1$

Figure 1. Local minimum geometries of four isomers of $C_{24}$ calculated at B3lYP/PC1 level. The C atoms are shown in grey

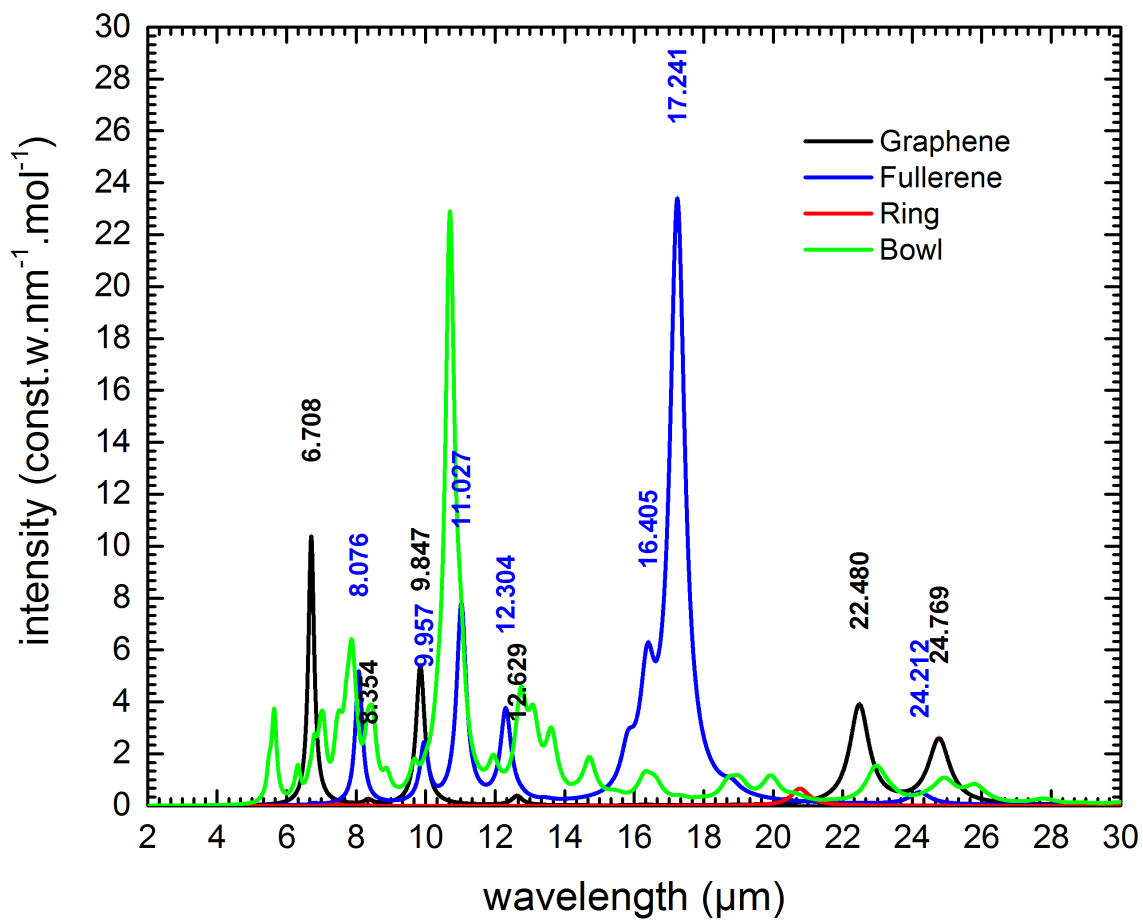

Figure 2. Simulated infrared emission spectra (*T*=500 K) of four isomers of $C_{24}$ at B3LYP/PC1 level

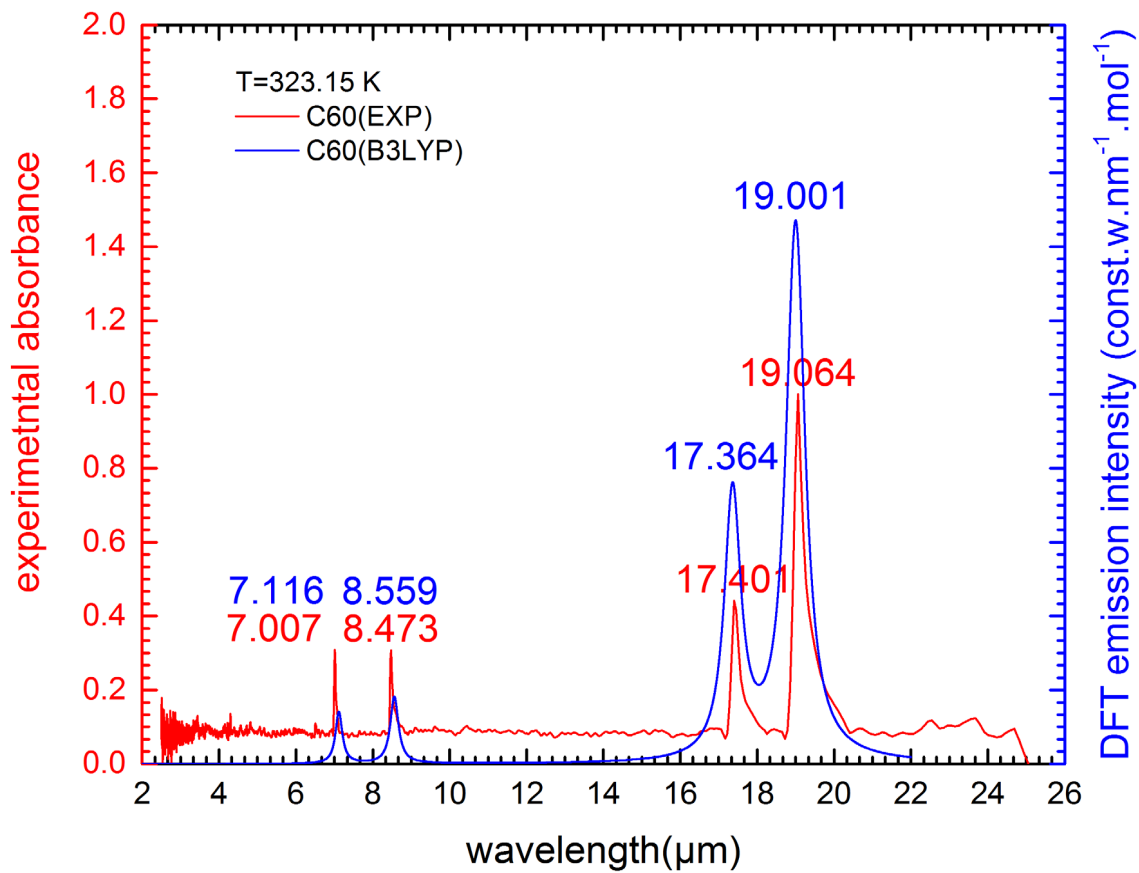

Figure 3. A comparison between solid state FTIR data and combined Drude-B3LYP/PC1 simulation at $T$=323.15 K for $C_{60}$.

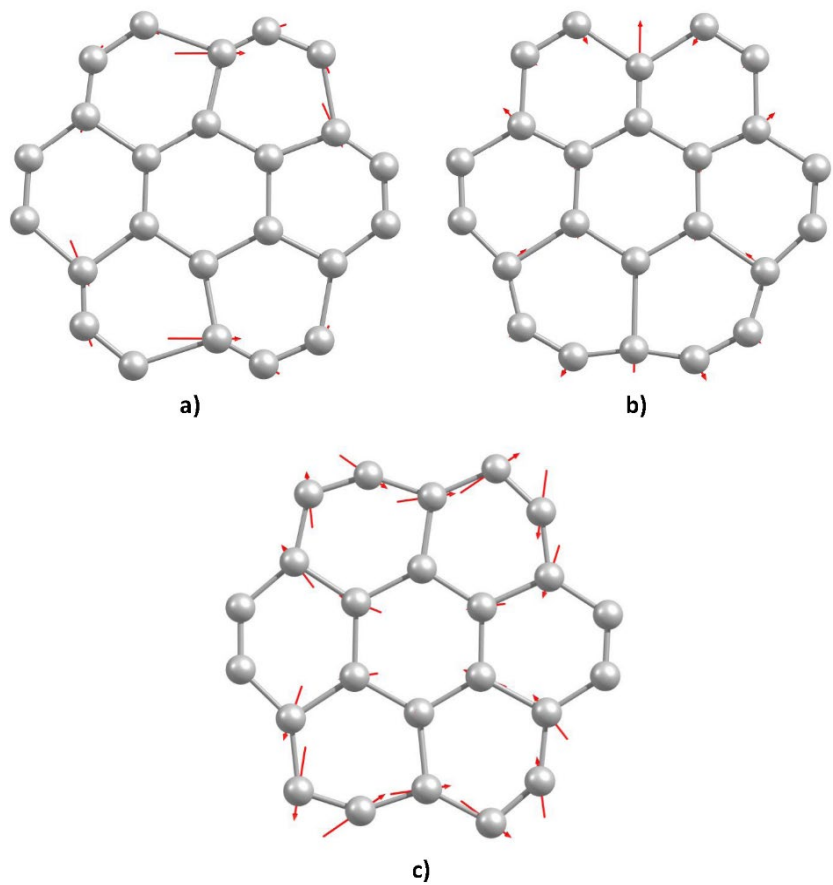

Figure 4. Vibrational motions in graphene form of $C_{24}$ at a) 6.708 μm, b) 9.847 μm and c) 22.480 μm